\documentstyle[multicol,aps,pre,epsfig]{revtex}

\begin{document}
\draft

\title{Dependence of heat transport on the strength and 
shear rate of prescribed circulating flows}

\author{Emily S. C. Ching and K. M. Pang}
\address{Department of Physics, The Chinese University of Hong Kong,
Shatin, Hong Kong}

\date{\today}

\maketitle

\begin{abstract}

We study numerically the dependence of heat transport 
on the maximum velocity
and shear rate of physical circulating flows, which are
prescribed to have the key characteristics of the large-scale mean 
flow observed in turbulent convection. 
When the side-boundary thermal layer is thinner 
than the viscous boundary layer, the Nusselt number (Nu), 
which measures the heat transport,
scales with the normalized shear rate to an exponent 1/3. On 
the other hand, when the side-boundary thermal layer is
thicker, the dependence of Nu on the Peclet number, which measures
the maximum velocity, or the normalized shear rate 
when the viscous boundary layer thickness is fixed, is
generally not a power law. Scaling behavior is obtained only in
an asymptotic regime. The relevance of our results to the problem of
heat transport in turbulent convection is also discussed.

\end{abstract}

\vspace{5pt}


\begin{multicols}{2}

\section{Introduction}

Turbulent Rayleigh-B{\'e}nard convection has been a system
of much research interest. The system consists of a closed
cell of fluid which is heated from below and cooled from above.
When the applied temperature difference is large enough,
the fluid moves and convection occurs. The flow state is
characterized by the geometry of the cell and two dimensionless 
control parameters: the Rayleigh number Ra, which measures how much
the fluid is driven and the Prandtl number Pr, which is the
ratio of the diffusivities of momentum and heat of the fluid.
The two parameters are defined by
Ra $= \alpha g \Delta L^3/ (\nu \kappa)$, and
Pr $= \nu/\kappa$, where $\Delta$ is the maintained temperature
difference, $L$ is the height of the cell,
$g$ the acceleration due to gravity, and $\alpha$,
$\nu$, and $\kappa$ are respectively the volume expansion coefficient,
kinematic viscosity and thermal diffusivity of the fluid.
When Ra is sufficiently large, the convection becomes turbulent. 

Besides the issue of the statistical characteristics of the
velocity and temperature fluctuations, it is of
interest to understand the heat transport by the fluid,
which is an overall response of the system.
The heat transport is usually expressed as the       
dimensionless Nusselt number Nu, which is the ratio of the measured
heat flux to the heat transported were there only conduction. Before
the onset of convection, heat is transported only by conduction and
Nu is identically equal to one. When convection occurs, heat is
more effectively transported by the fluid due to its motion and
Nu increases from 1. A major question is then to understand how
Nu depends on Ra and Pr.

The work of Libchaber and coworkers on turbulent convection in 
low temperature helium gas\cite{Libchaber1,Libchaber2} 
showed that Nu has a simple power-law dependence on Ra:
\begin{equation}
{\rm Nu} \sim {\rm Ra}^\beta
\label{NuRa}
\end{equation}
and the exponent $\beta$ is almost equal to $2/7$, which is 
different from $1/3$, the value that marginal stability 
arguments\cite{Malkus} would give. This result led to the 
development of several theories\cite{Libchaber2,She,SS,Cioni}
which all give $\beta = 2/7$ 
but were based on rather different physical assumptions. 
In particular, the Chicago mixing-zone model
\cite{Libchaber2} emphasizes the heat transport by the thermal plumes,
the coherent structures observed in turbulent convection while
the theory by Shraiman and Siggia\cite{SS} focuses on the effect of 
the shear of the large-scale mean flow on the heat transport 
(see e.g. Ref.\cite{review} for a review). 
Recent experimental results appeared to further
complicate the situation. Niemela et al.\cite{Niemela} reported
a value of $\beta$ close to 0.31 for
measurements in low temperature helium gas that cover 
a much larger range of Ra, from $10^6$ to $10^{17}$.
Xu et al.\cite{Xu} studied turbulent convection in acetone in several 
experimental cells of different aspect ratios and concluded that 
there is no significant range of Ra over which the scaling behavior 
Eq. (\ref{NuRa}) holds. Furthermore, they showed that for Ra $\ge 10^7$,
the dependence of Nu on Ra can be represented by a combination of
two power laws, which is consistent with a recent theory by Grossmann
and Lohse\cite{Lohse1,Lohse2}. In Grossmann and Lohse's theory,
the viscous and thermal dissipation were decomposed 
into their bulk and boundary-layer contributions, and ten
asymptotic regimes were obtained\cite{Lohse2}.

Another interesting feature observed in turbulent convection is the
presence of a persistent large-scale mean flow which spans the
whole experimental cell\cite{lsc}. The maximum mean velocity
of the flow was also found to scale as Ra to about 1/2\cite{Wu}. 
The presence of a large-scale flow
naturally induces an interaction between the top and bottom thermal
boundary layers. Such an interaction was taken to be absent in the marginal
stability arguments. One obvious effect of the velocity field, 
which satisfies the no-slip boundary condition, is that it produces a
shear near the boundaries, which was first studied in Ref.\cite{SS}.

In convection, the equations of motion are:
\begin{mathletters}
\begin{eqnarray}
&{\displaystyle \partial {\vec u} \over \displaystyle \partial t} 
+ {\vec u} \cdot {\vec \nabla} {\vec u}  
= -{\vec \nabla} p + \nu \nabla^2 {\vec u} + g \alpha T \hat z  \\
&{\displaystyle \partial T \over \displaystyle \partial t} 
+ {\vec u} \cdot {\vec \nabla} T = \kappa \nabla^2 T  \\
&{\vec \nabla} \cdot {\vec u} = 0 
\end{eqnarray}
\end{mathletters}where ${\vec u}$ is the velocity field, $p$ the pressure 
divided by density and $T$ the temperature field, while $\hat z$ is the 
unit vector in the vertical direction. The velocity and temperature are 
thus coupled dynamically in a complicated fashion 
and have to be solved together. Physically,
the velocity field is driven by the applied temperature difference. 
The flow in turn determines the temperature profile, 
and thus the heat transport, in a self-consistent manner.

To gain insights of the problem of
heat transport in turbulent convection, we have
turned to the simpler problem of heat transport by 
prescribed velocity fields that have features of the  
large-scale mean flow observed. The large-scale flow has two
dominant features: (i) it is a circulating flow that spans the whole
experimental cell and (ii) it generates a shear near the boundaries.
In an earlier paper, Ching and Lo studied separately 
these two features and their effects on
the heat transport\cite{ChingLo}. They found that Nu scales with the
Peclet number that measures the maximum velocity to an exponent 1/2 for
a purely circulating flow and scales with the normalized
shear rate to an exponent 1/3 for a pure shear. 

Pure circulating or pure shear flows are, however, not physical 
fluid flows within a closed box. 
In this paper, we continue
along these lines of thought and study the heat transport 
by {\it physical} velocity fields in a 
unit square cell that have both the
circulating and the shear-generating features.
We focus on the dependence of the heat transport 
on the maximum velocity and the 
shear rate of the flows. We first formulate our problem in Sec. II.
Then we present our numerical results and discuss how these results
can be understood in Sec. III. In Sec. IV, we further discuss how these
results are relevant to the understanding of heat
transport in turbulent convection. Finally, we end the paper with a
summary and conclusions in Sec. V.

\section{The Problem}

We solve the steady-state
advection-diffusion equation 
\begin{equation}
{\vec u}(x,y) \cdot \vec{\nabla} T(x,y) =  \kappa \nabla^2 T(x,y)  \ 
\label{adv-diff}
\end{equation}
for a prescribed incompressible velocity field ${\vec u}(x,y)$ in 
a unit square cell: $0 \le x \le L$ and $0 \le y \le L$. 
A temperature difference of $\Delta$ is applied
across the $y$-direction while no heat conduction is allowed
across the $x$-direction. That is, the temperature field
$T(x,y)$ satisfies the following boundary conditions:
\begin{eqnarray}
T(x,y=0) = \Delta;  &\qquad& T(x,y=L) = 0
\label{top_bottom_boundaries} \\
{\partial T \over \partial x}(x=0,y) &=&
{\partial T \over \partial x} (x=L,y) = 0
\label{side_boundaries}
\end{eqnarray}

\begin{figure}
\begin{center}
\mbox{\epsfig{file=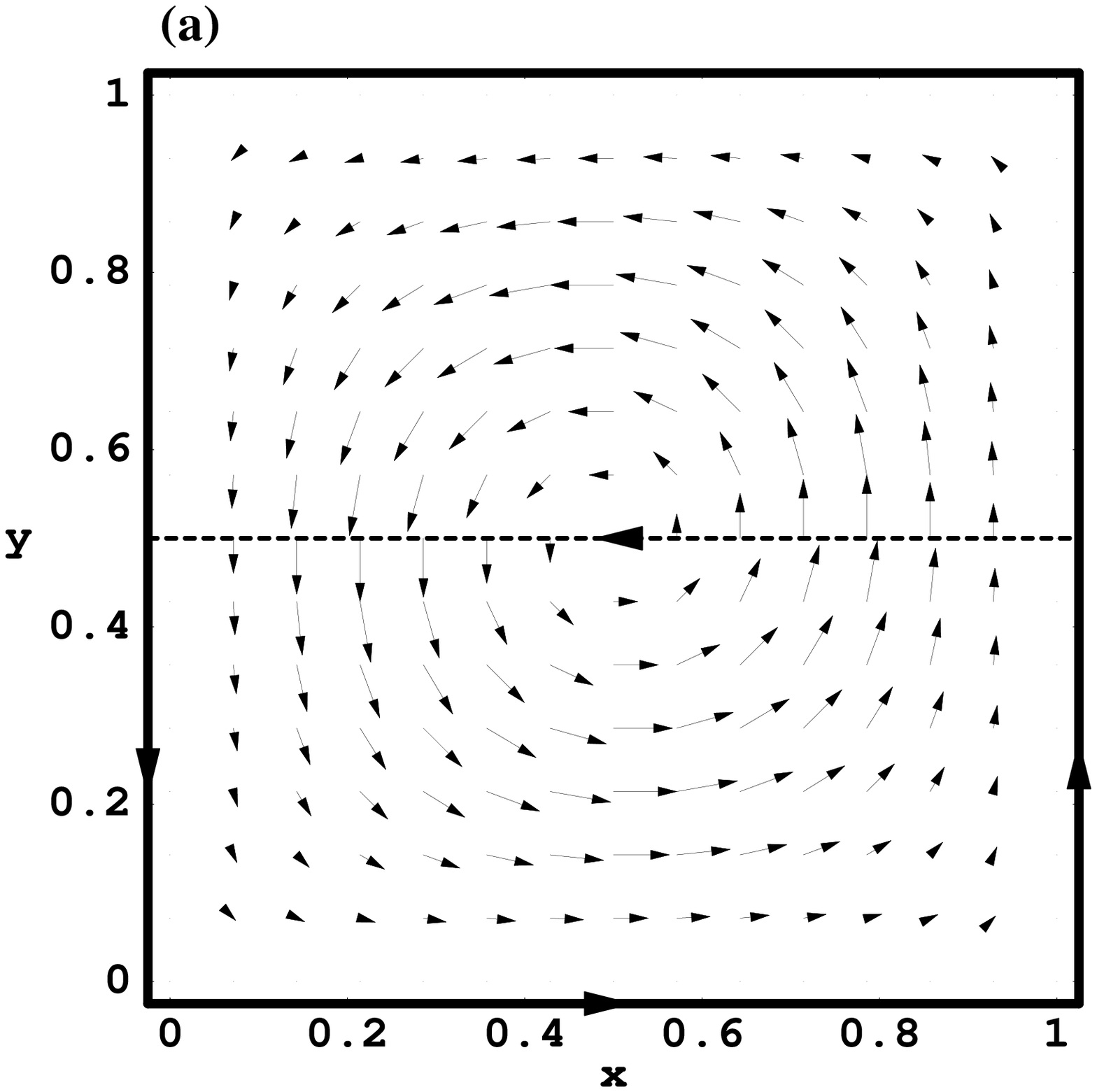,width=6.cm,angle=270}}
\mbox{\epsfig{file=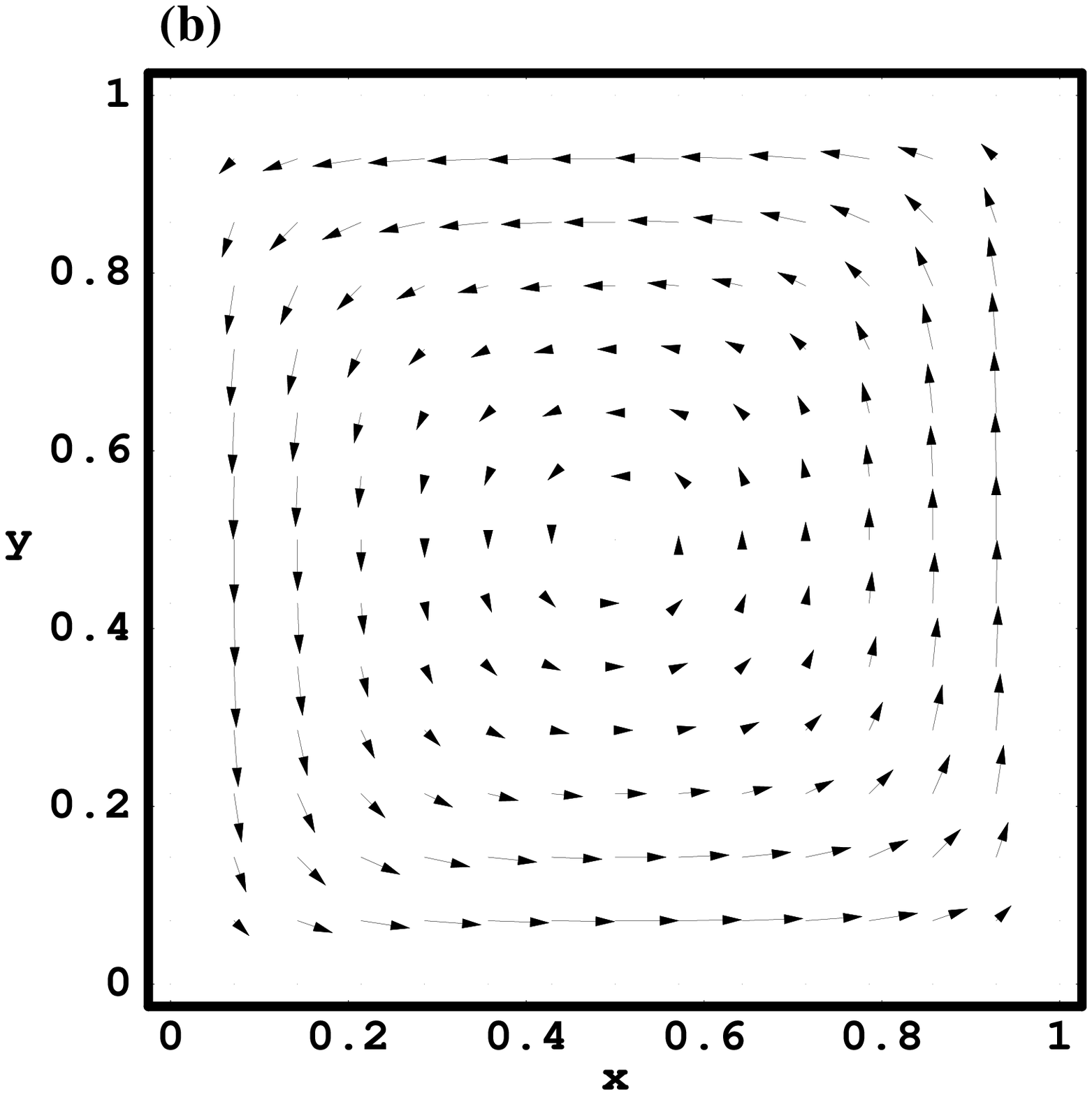,width=6.cm,angle=270}}
\vspace{-2cm}
\narrowtext
\caption{
(a) The velocity field with an alegbraic $f$ given by
Eq.~(\ref{algebraic}) with $a=b=1$.
The size of the arrow indicates the relative
magnitude of the velocity. The contour $C$ that encloses the lower
half of the unit square cell is also shown. (b) The velocity field
with an exponential $f$ given by Eq.~(\ref{exponential}) with $c=1$ and
$k=9$.}
\label{fig1}
\end{center}
\end{figure}

For the velocity field, we take 
\begin{eqnarray}
u_x(x,y) &=& f(\tilde x) f'(\tilde y) 
\label{ux}\\
u_y(x,y) &=& -f'( \tilde x) f(\tilde y)
\label{uy}
\end{eqnarray}
where $f$ is some function of $\tilde x \equiv x/L$ or 
$\tilde y \equiv y/L$ and $'$ is its derivative with
respect to the argument. Thus, $\vec{u}$ is incompressible
and separable in $x$ and $y$.
To satisfy the no-slip boundary condition, we require
\begin{equation}
f(0)=f(1)=f'(0)=f'(1)=0
\label{noslip}
\end{equation}
Moreover, for $\vec{u}$ to be a circulating flow, we require
\begin{equation}
f(\tilde x) = f(1-\tilde x) \qquad 0 \le 
\tilde x \le 1
\label{antisym}
\end{equation}
such that $u_x$ and $u_y$ are antisymmetric about
$y=L/2$ and $x=L/2$ respectively.
We have studied two forms of $f$. The first form is algebraic.
When $f$ is algebraic, Eqs. (\ref{noslip}) and (\ref{antisym})
imply that $f(\tilde x) = {\tilde x}^2({\tilde x}-1)^2 
h(\tilde x)$ with $h(\tilde x)=h(1-\tilde x)$.
Thus we choose
\begin{equation}
f(\tilde x) = {\tilde x}^2({\tilde x}-1)^2 
(a \tilde x+b)[a(1-\tilde x)+b]
\label{algebraic}
\end{equation}
where $a$ and $b$ are positive constants.
Equation (\ref{algebraic}) is the simplest algebraic form that allows
us to change both the circulating strength and the shear rate of the
flow (see below). The other is an exponential form of $f$:
\begin{equation}
f(\tilde x) = c(1-e^{-k \tilde x})^2[1-e^{-k(1-\tilde x)}]^2
\label{exponential}
\end{equation}
where $c$ and $k$ are positive constants. 
For $k \ll 1$, the exponential form reduces to the algebraic form with
$a=0$ and $b^2 = ck^4$. When $k$ is large, the velocity decays 
exponentially towards the center of the cell.
We show the two velocity fields in Fig.~\ref{fig1}.

We characterize the velocity field by its maximum velocity $u_0$ 
and shear rate $\gamma$, which are defined by
\begin{eqnarray}
u_0 &\equiv& \max_{0\le y \le {L \over 2}} \, u_x(x={L \over 2},y) =
\max_{0\le x\le {L \over 2}} \, -u_y(x,y={L \over 2} )
\label{u0} \\
\gamma &\equiv& 
{\partial u_x \over \partial y}(x={L \over 2},y) \bigg|_{y=0}
= -{\partial u_y \over \partial x}(x,y={L \over 2}) \bigg|_{x=0}
\label{gamma}
\end{eqnarray}
By varying the parameters $a$ and $b$ or $c$ and $k$, we can vary 
$u_0$ and $\gamma$ of the velocity fields. 

From the solved temperature field, we can calculate the heat
transport, which is the heat conducted across the boundary $y=0$.
Nu is therefore the ratio of the average of the magnitude of the
vertical temperature gradient
over the boundary $y=0$ to
$\Delta/L$:
\begin{equation}
{\rm Nu} = {\displaystyle
 \langle - {\partial T \over \partial y} \bigg|_{y=0} \rangle
\over \displaystyle {\Delta  \over L}}
\label{Nu}
\end{equation}
Here $\langle \ldots \rangle$ is the average over $x$ from 0 to $L$.
Our goal is to study the dependence of Nu on 
the Peclet number Pe $\equiv u_0 L /\kappa$
and the normalized shear rate $\tilde \gamma \equiv \gamma L^2/\kappa$. 

\section{Results and Discussions}

\begin{figure}[]
 \begin{center}
\mbox{\epsfig{file=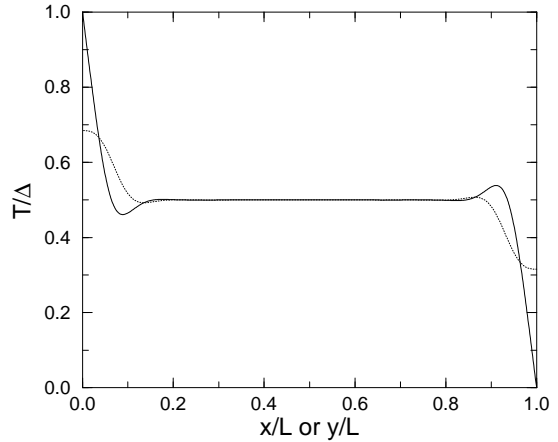,width=6cm,angle=270}}
\narrowtext
\caption{Typical vertical and horizontal temperature profiles
$T(x=L/2,y)/\Delta$~(solid) and $T(L-x,y=L/2)/\Delta$~(dotted)
as a function of $y/L$ and $x/L$ respectively.}
\label{fig2}
\end{center}
\end{figure}

\begin{figure}[]
 \begin{center}
\mbox{\epsfig{file=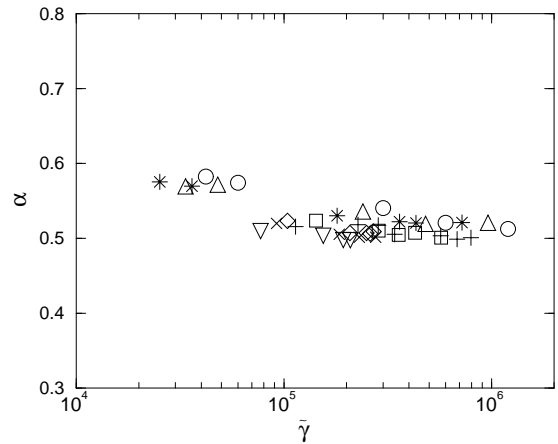,width=6cm,angle=270}}
\narrowtext
\caption{The dependence of $\alpha$ on $\tilde \gamma$
for various values of $\lambda_v$: $\lambda_v/L = 0.0167$ (circles),
$\lambda_v/L=0.021$~(triangles), $\lambda_v/L=0.028$~(stars),
$\lambda_v/L=0.070$~(squares), $\lambda_v/L=0.088$~(plusses),
$\lambda_v/L=0.096$~(diamonds), $\lambda_v/L=0.108$~(crosses),
and $\lambda_v/L=0.130$~(inverted triangles).}
\label{fig3}
\end{center}
\end{figure}

We numerically solve $T(x,y)$ for the two forms of $f$.
In Fig.~\ref{fig2}, we show the vertical and horizontal temperature 
profiles $T(x=L/2,y)$ and $T(L-x,y=L/2)$. As was reported in
turbulent convection experiments, the applied temperature difference
concentrates in two narrow regions near the `bottom' and `top'
boundaries $y=0$ and $y=L$ respectively. 
Interestingly, the circulation also induces a temperature
difference between the two `side' boundaries $x=0$ and $x=L$.
For flows that are anticlockwise~(see Fig.~\ref{fig1}), the side $x=0$
is colder than the side $x=L$.
The horizontal temperature profile $T(L-x,y=L/2)$ resembles
the vertical one $T(x=L/2,y)$ in that it is almost constant = 
$\Delta/2$ except for two small regions near the sides. In these
two small regions, it is approximately linear in $x$ except 
at the boundaries where the horizontal gradient vanishes. Hence,
we approximate $T(x,y=L/2)$ by
\begin{equation}
{T(x,y=L/2) \over \Delta/2 } \approx
\cases{1- \alpha \left(1- {x\over \ell}\right) & $0 \le x \le \ell$ \cr 
1 & $\ell < x < L- \ell$ \cr
1+ \alpha \left(1-{L-x \over \ell}\right) & $L -\ell \le x \le L$ }
\label{Tx}
\end{equation}
where $\alpha$ is about $0.5$ as shown in Fig.~\ref{fig3}.

\begin{figure}[]
 \begin{center}
\mbox{\epsfig{file=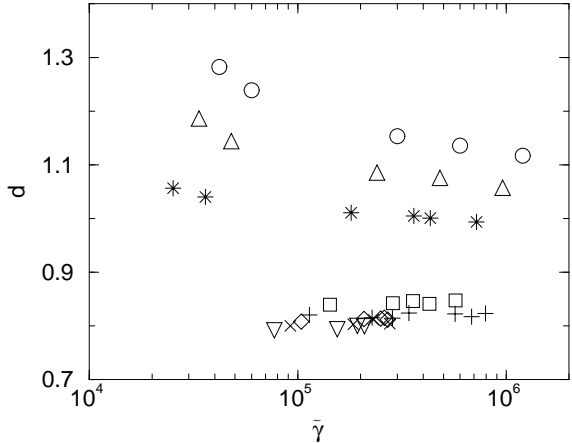,width=6cm,angle=270}}
\narrowtext
\caption{The dependence of $d$ on $\tilde \gamma$
for various values of $\lambda_v$~(same symbols as in Fig.~3).}
\label{fig4}
\end{center}
\end{figure}

Thus, there are also two thermal boundary layers of thickness $\ell$
at the `side' boundaries. Furthermore, we find that Nu is given by
$L/\ell$ up to a factor $d$:
\begin{equation}
{\rm Nu} = d {L \over \ell}
\label{d}
\end{equation}
where $d$ is weakly independent of $\tilde \gamma$ and approaches
$0.8$
as the viscous boundary layer thickness $\lambda_v \equiv
u_0/\gamma$
increases to 0.13~$L$~as shown in Fig.~\ref{fig4}.
Defining $\lambda_T= L/(2{\rm Nu})$ to be the average thickness of
the
thermal boundary layer at the top and bottom boundaries, we see that
the $\ell = 2d \lambda_T$. Thus, the side-wall thermal boundary
layers
are thicker than the top and bottom thermal boundary layers.

\begin{figure}[]
 \begin{center}
\mbox{\epsfig{file=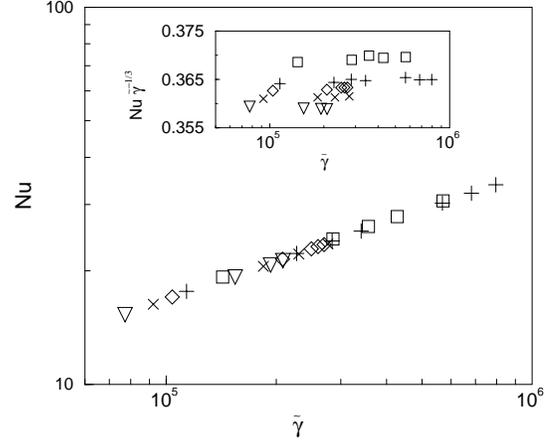,width=6cm,angle=270}}
\narrowtext
\caption{The dependence of Nu on $\tilde \gamma$ when $\ell <
\lambda_v$ for various values of $\lambda_v$ (same symbols as in
Fig.~\ref{fig3}).
In the inset, Nu${\tilde \gamma}^{-1/3}$ is plotted versus $\tilde
\gamma$.}
\label{fig5}
\end{center}
\end{figure}

\begin{figure}[]
 \begin{center}
\mbox{\epsfig{file=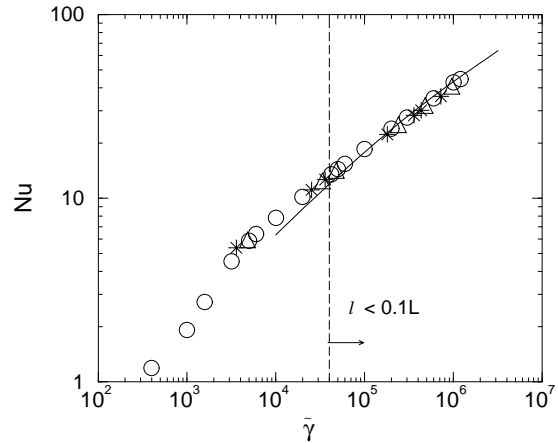,width=6cm,angle=270}}
\narrowtext
\caption{The dependence of Nu on $\tilde \gamma$ when
$L/2 \gg \ell > \lambda_v$ for various values of $\lambda_v$ (same
symbols as in Fig.~\ref{fig3}).
We also compare our numerical results to Eq.~(\ref{Nu2})
with $\lambda_v/L=1/60$, $\alpha=0.6$ and $d=0.9$ (solid line),
and good agreement is found for $\ell < 0.1L$ (data points on the
right
of the dashed line).}
\label{fig6}
\end{center}
\end{figure}

We find that the functional form of Nu(Pe,$\tilde \gamma$)
depends crucially on the relative sizes of
$\lambda_v $ and $\ell$.
For $\ell < \lambda_v$, Nu scales with the shear rate:
\begin{equation}
{\rm Nu} = A {\tilde \gamma}^{1/3}
\qquad {\rm for} \ \ell< \lambda_v  
\label{result1}
\end{equation}
and the coefficient $A$ approaches 0.36 as $\lambda_v$ increases to 
0.13~$L$ as shown in Fig.~\ref{fig5}. 
For $\lambda_v < \ell$, we see in Fig.~\ref{fig6} 
that the dependence of Nu on
Pe or $\tilde \gamma$ for fixed $\lambda_v$ is not a power law.
For a short range of Pe or $\tilde \gamma$, the dependence might be
described by an effective power law but the value of the effective 
exponent would depend on the range of Pe or $\tilde \gamma$ fitted.

We shall understand these results in the following.
Since the velocity field is incompressible, 
Eq. (\ref{adv-diff}) implies
\begin{equation}
\int_{C} 
\left( \vec{u} T - \kappa \vec \nabla T \right) \cdot \hat{n} d l = 0
\label{surface}
\end{equation}
for any closed curve $C$ in the two-dimensional domain where
$\hat{n}$ is an outward normal. Since we find that
$|\partial T/\partial y|$ almost vanishes along $y=L/2$,
we choose $C$ to enclose the lower half of the unit cell~(see
Fig.~\ref{fig1}a). Together with Eq. (\ref{side_boundaries}) and the
no-slip boundary condition, Eq. (\ref{surface}) then implies that
\begin{equation}
{\rm Nu} \approx {L \over \kappa \Delta} 
\langle u_y(x,y={L \over 2}) T(x,y={L \over 2}) \rangle 
\label{balance}
\end{equation}
Thus, we can estimate Nu by the heat transported across $y=L/2$.
Using the antisymmetry of $u_y$ about $x=L/2$
and Eq. (\ref{Tx}), we get
\begin{eqnarray}
\nonumber
& & \langle u_y(x,y={L \over 2}) T(x,y={L \over 2}) \rangle \\ 
\nonumber
&=& {2 \over L} \int_0^\ell u_y(x,y={L \over 2}) 
\left[T(x,y={L \over  2}) - {\Delta \over 2}\right] dx \\
&\approx& -{\alpha \Delta  \over L}
\int_0^\ell (1- {x \over \ell}) u_y(x,y={L \over 2}) dx
\label{est}
\end{eqnarray}
In the derivation of Eq.~(\ref{est}), we have assumed that the parameters
of the velocity field are chosen such that $\ell < L/2$ as in physical
situations. The heat transported across $y=L/2$ is, therefore, contributed
mainly by two `jets' of colder and hotter fluids moving down and up
respectively along the two boundaries $x=0$ and $x=L$.

\begin{figure}[]
 \begin{center}
\mbox{\epsfig{file=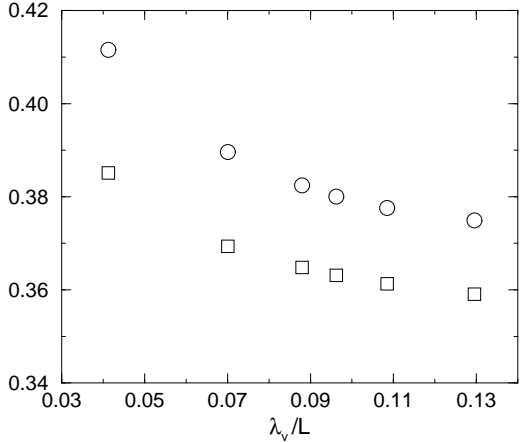,width=6cm,angle=270}}
\narrowtext
\caption{Comparison of the estimated values of the coefficient
$(\alpha d^2/6)^{1/3}$~(circles) with the computed values of
$A$~(squares).}
\label{fig7}
\end{center}
\end{figure}

Using Eq. (\ref{gamma}), we approximate
$u_y$ by a linear function in $x$ for $x < \lambda_v$: 
\begin{equation}
u_y(x,{L \over 2}) = -\gamma x \qquad {\rm for} \ x < \lambda_v
\label{linear}
\end{equation}
Hence,  Eqs. (\ref{d}), (\ref{balance}) and (\ref{est}) give
\begin{equation}
{\rm Nu} \approx  \left({\alpha d^2 \over 6}\right)^{1/3}
{\tilde \gamma}^{1/3} \qquad {\rm for} \ \ell< \lambda_v 
\label{Nu1}
\end{equation}
In Fig.~\ref{fig7}, we compare our estimated values of the coefficient 
$(\alpha d^2/6)^{1/3}$ with the computed values of $A$. It can
be seen that the two values agree within an error of 7\%.
Moreover, the agreement is better for larger $\lambda_v$,
as expected since the linear approximation Eq.~(\ref{linear}) 
works better for $0 < x \le \ell$ for larger $\lambda_v$.

For $\ell > \lambda_v$, we need $u_y$
beyond the region where a linear approximation holds. 
As a first approximation, we take
\begin{equation}
-u_y(x,y={L \over 2}) \approx \cases{\gamma x & $x < \lambda_v$ \cr
u_0 & $\lambda_v \le x \le \ell$ }
\label{uy_gen}
\end{equation}
That is, we approximate the large-scale flow by a shear near 
the boundaries then followed by a band of circulation at the maximum 
velocity. As $u_y$ has to decay to zero towards the center of the cell,
we expect the approximation Eq.~(\ref{uy_gen}) to work only for 
$\lambda_v < \ell \ll L/2$.

Using Eqs. (\ref{balance}), (\ref{est}) and (\ref{uy_gen}),
we get a quadratic equation for Nu.
Solving which gives
\begin{equation}
{\rm Nu} = { \sqrt{2 \alpha d {\rm Pe} - 
(\alpha^2/12){\rm Pe}^4 {\tilde \gamma}^{-2} }
- (\alpha/2) {\rm Pe}^2  {\tilde \gamma}^{-1} \over 
2\left(1-{\alpha \over 6d} {\rm Pe}^3 {\tilde \gamma}^{-2} \right) }
\label{Nu2}
\end{equation}
for $L/2 \gg \ell > \lambda_v$.
Thus, Nu does not have a power-law dependence on Pe nor $\tilde
\gamma$. Physically, this is because we generally cannot neglect 
the effect of the shear even when the edge of the side-wall thermal 
boundary layer is located at the band of maximum velocity of the 
large-scale flow. It is only in the limit Pe$^{3/2}/\tilde \gamma \ll 1$
that Nu $\sim$ Pe$^{1/2}$. Away from this asymptotic regime, Nu might
be represented by an effective power law on Pe or $\tilde \gamma$
for fixed $\lambda_v$ but the value of the effective exponent
would depend on the range of Pe or  $\tilde \gamma$ fitted.
In Fig.~\ref{fig6}, we compare the numerical results to
Eq.~(\ref{Nu2}) with $\lambda_v/L =1/60$, $\alpha = 0.6$ and $d = 0.9$. 
The agreement is not too bad for $\lambda_v < \ell < 0.1 L$
 given that our approximation of $u_y$ being 
constant beyond the shear layer is rather crude.

\section{Relevance to Turbulent Convection}

In turbulent convection, the heat transport is due both to the
large-scale mean flow and the fluctuating part of the velocity field.
Our present work provides insights only to the heat transport by
the large-scale mean flow. It is illuminating to see what results 
would be inferred if we neglect the effect of the fluctuating part of 
the velocity field.

Depending on the type of the viscous boundary layer, the strength and
the shear rate would be related to each other. For example, if we
follow Grossmann and Lohse to assume that
viscous boundary layer is of Blasius type\cite{LL}:
$\lambda_v/L \sim \sqrt{{\rm Pe}/{\rm Pr}}$
for moderate values of Pr\cite{Lohse1} and that $\lambda_v \sim L$ for
very large values of Pr\cite{Lohse2}, then
\begin{equation}
\tilde \gamma \sim \cases{{\rm Pe}^{3/2} {\rm Pr}^{-1/2}& {\rm
moderate \ Pr} \cr
{\rm Pe} & {\rm very \ large  \ Pr}}
\label{gammaPe}
\end{equation}
Our results Eqs.~(\ref{Nu1}) and (\ref{Nu2}) would thus
give Nu as a function of Pe and Pr:
\begin{equation}
{\rm Nu} \approx \cases{\left({\alpha d^2 \over 6 }\right)^{1/3} 
{\rm Pe}^{1/2} {\rm Pr}^{-1/6} & {\rm moderate \ Pr} \cr
\left({\alpha d^2 \over 6 }\right)^{1/3}
{\rm Pe}^{1/3} & {\rm very \ large \ Pr}}
\label{1}
\end{equation}
for $\ell < \lambda_v$
and 
\begin{equation}
{\rm Nu} \approx g({\rm Pr}) {\rm Pe}^{1/2}
\label{2}
\end{equation}
for $L/2 \gg \ell > \lambda_v$ with
\begin{equation}
g({\rm Pr})={\sqrt{\alpha d- \alpha^2 {\rm Pr}/12} -\alpha {\rm
Pr}^{1/2}/2 \over 2[1-\alpha {\rm Pr}/(6 d)]}
\end{equation}
We emphasize that the non-power-law dependence on Pr for 
$\ell > \lambda_v$ echoes that the effect of the shear
cannot generally be neglected even when the edge of the side-wall
thermal boundary layer is located at the band of the maximum velocity
of the large-scale flow.

Next, we make use of a rigorous relation between the viscous dissipation
and Nu and Ra, which is derivable from the equations
of motion\cite{SS,review}:
\begin{equation}
\langle \langle [\partial_i u_j({\bf x},t)]^2 \rangle \rangle
= {\kappa^2 \over L^4} ({\rm Nu}-1) {\rm Ra}
\label{exact}
\end{equation}
where $\langle \langle \ldots \rangle \rangle$ is an average
over space {\bf x} and time $t$.
Neglecting the contribution from the fluctuating part of the velocity
field to the viscous dissipation, we get
\begin{equation}
\tilde \gamma {\rm Pe} \sim {\rm Nu Ra}
\label{dom}
\end{equation}
for Nu $\gg 1$.
Using Eqs.~(\ref{gammaPe}), (\ref{1}), (\ref{2}) and (\ref{dom}),
we finally get
\begin{equation}
{\rm Nu} \sim {\rm Pr}^{-1/12} {\rm Ra}^{1/4}, \ \
{\rm Pe} \sim {\rm Pr}^{1/6} {\rm Ra}^{1/2} 
\label{moderatePr}
\end{equation}
\begin{equation}
{\rm Nu} \sim {\rm Ra}^{1/5}, \ \
{\rm Pe} \sim  {\rm Ra}^{3/5} 
\label{largePr}
\end{equation}
for $\ell < \lambda_v$ respectively for moderate and very large
values of Pr, and
\begin{eqnarray}
{\rm Nu} &\sim& g({\rm Pr})^{5/4}{\rm Pr}^{1/8} {\rm Ra}^{1/4} \cr
{\rm Pe} &\sim& g({\rm Pr})^{1/2} {\rm Pr}^{1/4} {\rm Ra}^{1/2} 
\label{large}
\end{eqnarray}
for $L/2 \gg \ell > \lambda_v$.
The results Eqs.~(\ref{moderatePr}), (\ref{largePr}), and (\ref{large}), 
except for the non-power-law dependence on Pr in 
Eq.~(\ref{large}), resemble those obtained by Grossmann and 
Lohse\cite{Lohse1}, respectively for regimes $I_l$, $I_\infty^<$,
and $I_u$, in which they took the boundary layer to dominate 
both the viscous and thermal dissipation.

\section{Summary and Conclusions}

In this paper, we have studied the heat transport by physical fluid flows 
whose velocity fields are prescribed to have both the two key 
characteristics of the large-scale mean flow observed in turbulent 
convection.  The velocity fields that we have chosen are separable and
incompressible circulating flows in a unit square cell. Satisfying the
no-slip boundary condition, they also generate a shear near the
boundaries. Overall, they are approximately a shear near the boundaries,
almost constant with the maximum strength of circulation for a finite
band, and then a decay towards the center of the cell.
We focus on the functional dependence of Nu on
Pe measuring the maximums strength of circulation 
and the normalized shear rate $\tilde \gamma$ that characterize
the velocity fields. 

We have shown that Nu can be estimated by the heat transported across 
the mid-height of the unit square cell, which is in turn contributed 
mainly by two jets of hotter and colder fluids moving up and down the
two sides. These two jets are confined to two narrow regions. That is,
the velocity field also induces two thermal
boundary layers at the side boundaries. These side-boundary
thermal layers are thicker than those at the top and bottom 
boundaries.
It is then clear that the functional form of
Nu(Pe,$\tilde \gamma$) depends crucially on the relative sizes
of the viscous boundary layer thickness $\lambda_v$
and the thickness of the thermal boundary layers at the
side boundaries $\ell$. When $\ell < \lambda_v$, that is, the edge
of the side-wall thermal boundary layer falls within
the shear region of the large-scale flow,
Nu scales with $\tilde \gamma$ to 1/3 and is very weakly dependent on
$\lambda_v$ or Pe. When $\ell > \lambda_v$ and the edge of the side-wall
thermal boundary layer falls within the band
of circulation with maximum strength, there is still contribution
to the heat transport by the shear and Nu depends on both
Pe and $\tilde \gamma$. The dependence is generally not
a power law and scaling behavior is obtained only in
the asymptotic regime Pe$^{3/2}/\tilde \gamma \ll 1$.

We have further discussed how our results are relevant to
the problem of heat transport in turbulent convection.
In turbulent convection, the heat transport is due both to
the large-scale mean flow and the fluctuating part of the
velocity field. Neglecting the effect of the fluctuating part 
of the velocity field, our results lead to results 
resembling those obtained by Grossmann and Lohse\cite{Lohse1,Lohse2} 
when they took the boundary layer 
to dominate both the viscous and thermal dissipation.
It is not surprising that the boundary layer dominating the 
viscous dissipation is the same as the 
large-scale mean flow dominating the viscous dissipation
since the gradient of the large-scale mean flow, which contributes to the
viscous dissipation, concentrates in the boundary. 
Our finding thus suggests that whether the boundary layer or the bulk
dominates the thermal dissipation is physically
equivalent to whether the large-scale mean flow or the fluctuating part
of the velocity field dominates the heat transport. 

\acknowledgements
We acknowledge D. Lohse for pointing out to us that we can also derive
Eq.~(\ref{largePr}) for very large Pr
and thank P.T. Leung for discussions.
This work is supported by a grant 
from the Research Grants Council of the Hong Kong Special 
Administrative Region, China (RGC Ref. No. CUHK 4119/98P).

\end{multicols}
\end{document}